\documentclass[cits]{PoS}
\usepackage[authoryear,square]{natbib}
\bibpunct{(}{)}{;}{a}{}{,}
\usepackage{upgreek}
\newcommand{\SKAMid}{\rm SKA-{MID}}

\let\ifpdf\relax

\title{A Cosmic Census of Radio Pulsars with the SKA}

\ShortTitle{Pulsar Census}

\author{\speaker{E.~F.~Keane}$^{1,2,3}$, B.~Bhattacharyya$^{4,5}$,
  M.~Kramer$^{6,4}$, B.~W.~Stappers$^{4}$, S.~D.~Bates$^{7}$,
  M.~Burgay$^{8}$, S.~Chatterjee$^{9}$, D.~J.~Champion$^{6}$,
  R.~P.~Eatough$^{6}$, J.~W.~T.~Hessels$^{10,11}$, G.~Janssen$^{10}$,
  K.~J.~Lee$^{12,6}$, J.~van~Leeuwen$^{10,11}$, J.~Margueron$^{13}$,
  M.~Oertel$^{14}$, A.~Possenti$^{8}$, S.~Ransom$^{15}$,
  G.~Theureau$^{16}$ \& P.~Torne$^{6}$ \\ 

\tiny{$^{1}$ Swinburne University of Technology, Australia;
$^{2}$ ARC Centre of Excellence for All-Sky Astrophysics (CAASTRO);
$^{3}$ E-mail: \email{ekeane@swin.edu.au};
$^{4}$ University of Manchester, UK;
$^{5}$ E-mail: \email{bhaswati.bhattacharyya@manchester.ac.uk};
$^{6}$ Max-Planck Institute for Radio Astronomy, Bonn, Germany;
$^{7}$ National Radio Astronomy Observatory, Green Bank, USA;
$^{8}$ INAF-Osservatorio Astronomico di Cagliari, Italy;
$^{9}$ Cornell University, USA;
$^{10}$ Netherlands Institute for Radio Astronomy (ASTRON), The Netherlands;
$^{11}$ University of Amsterdam, The Netherlands;
$^{12}$ Peking University, P.R.China;
$^{13}$ Institut de Physique Nucl\'{e}aire de Lyon, CNRS, France;
$^{14}$ Laboratoire Univers et Th\'{e}ories, Paris Observatory, France;
$^{15}$ National Radio Astronomy Observatory, Charlottesville, USA;
$^{16}$ Laboratoire de Physique et Chimie de l'Environnement et de l'Espace, CNRS-Universit\'{e} d'Orl\'{e}ans, France.
}
}

\abstract{The Square Kilometre Array (SKA) will make ground breaking
  discoveries in pulsar science. The wide field-of-view, high
  sensitivity, multi-beaming and sub-arraying capabilities, coupled
  with advanced pulsar search backends, will result in the discovery
  of a large population of pulsars and new, high-quality data on
  select sources from the already known pulsar population. In this
  chapter we outline the surveys for new pulsars, as well as how we
  will perform the necessary follow-up timing observations of new
  discoveries. Pulsar surveys are essential to enable all of the SKA's
  headline pulsar science goals (tests of General Relativity with
  pulsar binary systems, investigating black hole theorems with
  pulsar-black hole binaries, and direct detection of gravitational
  waves in a pulsar timing array). Using SKA1-MID and SKA1-LOW at
  several different sky frequencies, we will survey the Milky Way to
  unprecedented depth, increasing the number of known pulsars by more
  than an order of magnitude. SKA2 will potentially enable us to find
  all of the Galactic radio-emitting pulsars in the SKA sky which are
  beamed in our direction. This will give us a clear picture of the
  birth properties of pulsars and of the gravitational potential,
  magnetic field structure and interstellar matter content of the
  Galaxy. Our targeted searches will enable detection of the most
  exotic systems, such as the $\sim1000$ pulsars we infer to be
  closely orbiting Sgr A*, the supermassive black hole in the Galactic
  Centre. In addition to Galactic pulsars, the sensitivity of the SKA
  will be sufficient to detect pulsars from local group galaxies; we
  can use these sources as probes of the local intergalactic
  medium. All of the discoveries will require regular 
  re-observations for a few months in order to derive their spin
  characteristics and establish the particular science questions they
  can be used to address. To do this efficiently we will perform live
  searches, and use sub-arraying and dynamic scheduling to time
  pulsars as soon as they are discovered, while simultaneously
  continuing survey observations. The large projected number of
  discoveries suggests that we will uncover currently unknown rare
  systems that can be exploited to push the boundaries of our
  understanding of astrophysics and provide tools for testing physics,
  as has been done by the pulsar community in the past.}

\FullConference{
Advancing Astrophysics with the Square Kilometre Array\\
June 8-13, 2014\\
Giardini Naxos, Italy}

\begin{document}

\section{Introduction}
The SKA will be a discovery machine. Apart from delivering
transformational science based on the expected huge increase in pulsar
timing precision, the SKA's high sensitivity, wide field of view
(FoV), and frequency coverage will allow us to explore the variable
radio sky in an unprecedented way. This will lead to the discovery of
previously unknown types of sources and enable us to probe a wide
range of explosive and dynamic events. Eventually, it will lead to a
full census of detectable radio pulsars in the Milky Way and
beyond. Among the new sources will be fast, spin-stable millisecond
pulsars (MSPs) whose period distribution reflects the equation of
state of nuclear matter and some of which will serve as detectors of
nano-Hz gravitational waves (GWs). Relativistic binary pulsars,
particularly those with orbital periods of a few hours or less, will
allow strong-field tests of General Relativity and other theories of
gravity. The discovered pulsars will also be superb probes for an
enhanced understanding of the Milky Way, its structure and its
constituents, including magnetic fields, the free electron
distribution etc. Other applications across a wide range of physics
and astrophysics topics are described in the accompanying chapters.

The first step toward these unique achievements will be enabled by
Phase I of the SKA (referred to as SKA1 hereafter). Apart from being a
transformational telescope in its own right, the first science phase
of this unique telescope will set the scene for the experiments to be
conducted with the full array. However, it is important to realize
that for searching, SKA1 is not only a simple stepping stone towards
SKA2. Due to limitations in processing power, it is unlikely that the
full area of the completed SKA can be utilized for a blind,
large-scale survey for some time to come. Hence, SKA1, with a highly
concentrated core, represents a significant fraction of the collecting
area usable for surveys with SKA2, and substantial achievements can be
made in pulsar searching in the early science phase.

\begin{figure}
  \centering
  \includegraphics[width=4in]{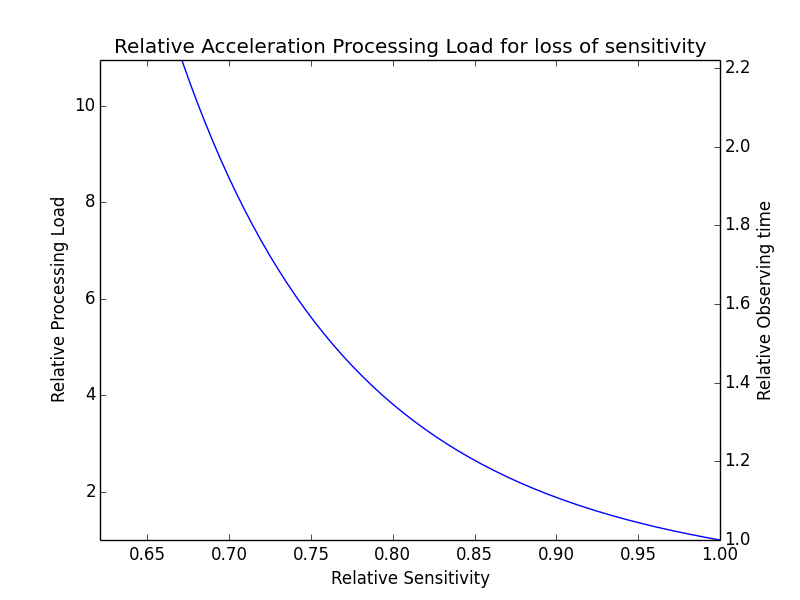}
  \caption{\footnotesize{The `cost', in terms of the increased
      observing time and (most significantly) computer processing
      required to find accelerated binary systems, in the event of a
      reduction in sensitivity.}}
\label{fig:cost}
\end{figure}

When considering the specifications for SKA1, we point out that, in
general, a loss in sensitivity (i.e. collecting area) cannot be simply
compensated by longer integration times. In the best case, a reduction
in collecting area that can be phased up coherently may require a
significant increase in computing power (both for beam-forming, and
processing). These costs are typically prohibitive, e.g. as
illustrated by Figure~\ref{fig:cost} a $30\%$ loss in raw sensitivity,
if compensated for by doubling the observing time, results in a
ten-fold increase in computation in order to find \textit{the same
  pulsars}. Rephrased, this shows that as SKA1 approaches the optimal
combination of dishes in the core region, pulsar searching becomes
ever more efficient. In the worst case, a loss in sensitivity means a
degradation in the science that is possible.

\subsection{Pulsar Searching Basics} 
The pulsar signal is periodic with known periods covering nearly four
orders of magnitude, i.e.~$1$~ms to $10$~s. The pulse duty cycle
ranges, typically, from less than one degree in rotational phase to
the full pulse period. Hence, 
the discovery of pulsars requires a search of the radio sky
in dispersion measure, pulse period and pulse duty cycle. For binary
pulsars, a search for acceleration is also required. The standard
search technique is consequently computationally expensive and it
involves the Fourier-transform of correspondingly prepared time-series
where the search in duty cycle uses a technique known as {\em harmonic
  summing}. Apart from these data processing requirements, forming
beams for a sufficiently large FoV to enable reasonable survey speed, 
is computationally challenging in the signal
processing chain, in particular if the array configuration is not
sufficiently compact. In order to ease the computational requirements,
we have explored different observing strategies for the search for
pulsars, depending on sky (i.e. Galactic) location, frequency (and
hence receiver technology), and observing modes. 

\subsection{Acceleration searches} 
Arguably, the most interesting systems to be found will be highly
accelerated. The SKA promises an advance over existing telescopes in
the particular ability to allow for short integration times due to its
large sensitivity. That means that only a small fraction of the orbit
is sampled, so that the pulse frequency may not change significantly
due to varying Doppler shifts. In contrast, reduced sensitivities
imply longer integration times, which mean more Doppler-smearing that
needs to be compensated for (if possible at all) by computational
means. In the best case, one can assume constant acceleration.
Even in the simplest acceleration search, the computing power needed
scales with the cube of the observing time. In other words, a
10\%-reduction in sensitivity implies a required increase in computing
efforts by a factor of two in order to find the same systems (see
Figure~\ref{fig:cost}). In practice, the penalty is much higher, for
two reasons. Firstly, the previous assumption of constant acceleration
will be wrong for the most compact systems when the integration time
becomes a significant fraction of the orbital period (depending also
on the unknown eccentricity of the system). Secondly, even if antennae
at longer baselines could be included coherently in effective
collecting area, it would require an increased amount of beam forming.

The impact on the population so-lost is difficult to quantify as the
result depends not only on the (unknown) luminosity of the sources,
but also on the orbital parameters. 
Population syntheses are used to infer information (see
\S~\ref{sec:sims}), but we can be certain that there are more compact
orbits than the Double Pulsar ($P_{\mathrm{b}}=147$~min) or the
shortest pulsar binaries known ($P_{\mathrm{b}}=92$~min). It is
notable that the Double Pulsar was not detected in the regular
pointings of the Parkes Multi-beam Pulsar Survey ($t_{\rm
  int}=35$~min) because of Doppler-smearing. Only in the 10-times
shorter integrations of the ``PH''-survey, was it finally
discovered. It is conceivable that most of the strong
(mildly-accelerated) sources are discovered with the currently
available sensitivity and computer resources. This would imply that
the remaining sources are either relatively weak or very highly
accelerated. In other words a loss in SKA sensitivity (the largest
that we can ever expect to achieve) would suggest a certain (and
final) blindness to the most exciting systems.

SKA radio pulsar surveys will produce a large volume of prospective
candidates, the majority of which will be forms of noise. Typically,
such large numbers of candidates need to be visually inspected in
order to determine if they are real pulsars. This process can be
labour intensive and delay the candidate confirmation. In order to
process pulsar searching candidates in a real time fashion, computer
software to perform candidate ranking or classifying is demanded. At
present, software using machine learning
\citep{eatough10,lyon2013,Zhu14} or statistical
classifiers~\citep{lee13,mbb+14} can increase the identification rate
by a factor of about 50 to 1000. This will significantly improve the
efficiency of pulsar searches with SKA1 and SKA2, allowing for fast
identification and confirmation of the candidates.
 
\subsection{Structure of this Chapter}
The remainder of this chapter is structured as follows: In
\S~\ref{sec:parameter_space} we describe the large parameter space
probed in pulsar searches. Surveys with SKA-LOW and SKA-MID are
described in \S~\ref{sec:skalow} and \S~\ref{sec:skamid},
respectively. \S~\ref{sec:sims} describes the results of population
syntheses, and the results of simulations of the survey yields, and
evaluates different search strategies. In addition to the `blind'
surveys we describe targeted searches in \S~\ref{sec:target}. We
review, in \S~\ref{sec:discoveries}, the wide range of anticipated
discoveries and their scientific importance. Finally, in
\S~\ref{sec:followup} we discuss the important issue of the follow-up
timing strategies of the newly discovered pulsars which are needed to
maximize the scientific yield.

\section{Pulsar searches with the SKA}

\subsection{Parameter-space for search}\label{sec:parameter_space}
As described above, and elsewhere in this book, one of the key aims of
the SKA, in both Phase 1 and with SKA2, is to discover as many pulsars
as possible. The pulsars of interest for the different astrophysical
goals may be located anywhere within, or even beyond, our Galaxy and
so it is necessary to perform a ``blind'' survey of the entire sky
visible from the sites in Australia and South Africa. To survey this
area efficiently we need to achieve the maximum possible survey speed
and this requires excellent sensitivity combined with a
wide-FoV~\citep{sks+09}. The high time resolution required to discover
pulsars means that it is not
effective to search for them in the images that are the traditional
data product of interferometers like the SKA. Instead we have to form
the coherent sum of as many dishes/stations as possible to give us the
required sensitivity. However, as the dishes are sparsely distributed,
the FoV shrinks as we add more dishes to improve our sensitivity. This
can be overcome by forming more beams, but that comes with a computing
cost associated with forming and processing those beams.
Depending on the exact configuration of the all-sky pulsar survey to
be undertaken with SKA1 between 1500 and 2200 tied-array beams are
required for SKA1-MID and 500 tied-array beams are required for
SKA1-LOW. With SKA2 at least 10,000 beams will be required, which will
allow more dishes to be included in the array, improving sensitivity,
but while also increasing the FoV (this increased `survey speed' also
allowing longer integrations).

The radio emission from pulsars is dispersed, before it arrives at
Earth, by the free electrons along the line of sight to the
pulsar. This causes a frequency-dependent delay which needs to be
removed in order to recover the pulsed signal. Unfortunately, the
degree of dispersion cannot be predicted and so a search over a range
of so-called dispersion measures is required. The maximum dispersion
measure is of course highly dependent on where the pulsar is located,
and with the majority of pulsars located in the Galactic plane, where
the dispersion can reach its highest values, it is necessary to search
over a wide range of dispersion measures. This is very important for
the SKA where the sensitivity is high enough to allow us to discover
the most distant sources.  With SKA1 it is proposed that the search be
possible out to dispersion measures of 3000 DM units; with the
proposed search parameters --- $|\mathrm{acc}|<350$~m/s/s, $2048$
beams and harmonic folding ---
the computaional requirement is at least 7 Petaflop/s. However, with
SKA2 and/or higher frequency observations it should be possible to
search out to dispersion measures of 10,000 DM units and beyond.

To search for the systems critical to the key gravity studies, such as
the pulsar-black hole (PSR-BH) and double neutron star (DNS) systems,
requires that one is able to at least partially correct the modulation
of the pulse frequency caused by its motion in the binary --- usually
via acceleration searches.
The orbital acceleration that might be observed in compact PSR-BH
binaries could be an order of magnitude larger than that seen in the
most relativistic DNS systems (see \S~\ref{sec:psrbh}). For the
proposed integration time of SKA1 (10~min, see \S~\ref{sec:skamid})
the aim is to search for accelerations $>1000$~m/s/s; indeed using
fluctuation frequency domain methods this will be readily achievable
for spin periods greater than $>16$~ms~\citep{lk05}. In highly
eccentric PSR-BH systems, the acceleration can be larger by factors of
a few near periastron. While at the moment acceleration searches (in
both the time and frequency domain) over such ranges and with uniform
spin frequency sensitivity are computationally prohibitive, for SKA
this should be the goal.

\subsection{An all-sky pulsar survey with SKA-LOW}\label{sec:skalow}





All-sky pulsar searches are often most efficiently done using low
radio observing frequencies ($100 - 600$\,MHz).  Such surveys benefit
from the typically steep spectra of pulsars ($S \propto \nu^{-1.6}$;
\citealt{bat+13}), as well as the naturally larger FoV of the
telescope (FoV scales as $\nu^{-2}$ for a fixed aperture size or
baseline length). Interstellar propagation effects, most notably
dispersive delay, multi-path propagation due to scattering, and sky
temperature, are the main limitations at low frequencies. Dispersive
delay, which scales as $\nu^{-2}$, is a correctable effect however, as
long as the channel bandwidth is sufficiently narrow that
intra-channel smearing of the pulse is minimal compared with the
sampling time. Scattering, which scales as $\sim \nu^{-4}$, is not
practically correctable in a blind survey and is a major limitation
for finding distant pulsars in the Galactic plane, especially at short
rotational periods.
Sky temperature scales as $\nu^{-2.6}$, but is only $\sim 35$\,K (at
400\,MHz) for high Galactic latitudes. For these reasons low-frequency
pulsar surveys are the preferred option for Galactic latitudes beyond
$|b| \approx 5$\,degrees.

Of the $\sim2300$ known pulsars: there are 766 pulsars with $|b| >
5$\,degrees; of the 172 Galactic Fields MSPs (P $<30$~ms): 113,
i.e. {\it two thirds} are at $|b| > 5$\,degrees~\citep{man+05}. 
Furthermore, population synthesis of the entire Galactic pulsar
population indicates that if SKA1-LOW and SKA1-MID are used in a
complementary way to survey the entire SKA-visible sky, then the
maximum yield of pulsar discoveries can be achieved (assuming a
constant on-sky survey time) if SKA1-LOW surveys at Galactic latitude
$|b| > 5$\,degrees and SKA1-MID surveys the Galactic plane.

There are currently three major, ongoing low-frequency pulsar surveys:
(i) the Arecibo 327-MHz Drift
Survey\footnote{http://www.naic.edu/~deneva/drift-search/}, operating
from $300-350$\,MHz; (ii) the Green Bank Northern Celestial Cap
Survey\footnote{http://arcc.phys.utb.edu/gbncc/}, which in a certain
sense is a continuation of the Green Bank Drift-Scan
Survey\footnote{http://astro.phys.wvu.edu/GBTdrift350/} as well as the
GBT350 Galactic Plane survey, and is operating from $300-400$\,MHz;
and (iii) the LOFAR Tied-Array All-Sky
Survey\footnote{http://www.astron.nl/lotaas/}, which is operating from
$119-151$\,MHz and is very similar in approach to how SKA-LOW will
survey for pulsars and fast transients. Together, these low-frequency
GBT, Arecibo and LOFAR surveys have discovered 176 pulsars in the last
decade. The GBT surveys have used scan lengths of $\sim 2$\,minutes
and reached a minimum flux density $S^{350}_{\mathrm{min}} \approx
1$\,mJy. Each pointing covers 0.25\,sq. degrees. The Arecibo survey
provides only 40-s in-beam time, and reaches $S^{350}_{\mathrm{min}}
\approx 0.5$\,mJy. Each pointing covers 0.05\,sq. degrees. LOFAR's
much larger FoV --- which is enabled by the use of 219 simultaneous
tied-array beams to cover 9\,sq. degrees of sky --- allows 1-hr
integrations, which reach $S^{150}_{\mathrm{min}} \approx 1$\,mJy.

SKA1-LOW will provide an enormous leap in sensitivity and survey
efficiency compared with these ongoing surveys. 
An all-sky survey using just the 600-m (radius) core of SKA1-LOW along
with 10-min dwell time and 500 tied-array beams will reach
$S^{350}_{\mathrm{min}} \approx 0.05$\,mJy (ten times deeper than any
ongoing survey) and will cover $\sim1$\,sq. degrees per pointing.
We also note that the instantaneous sensitivity of the 600-m SKA1-LOW
core at 350\,MHz will be a factor of a few greater than
Arecibo. Instantaneous sensitivity is the important factor for
detecting weak individual pulses, like those from the intermittent
radio pulsars and the fast radio bursts (see the Chapter on ``Fast
Transients at Cosmological Distances'', \citealt{mkg+14}, for a
discussion of this).
We emphasize
that a cosmic census of radio-emitting neutron stars requires the
complementarity provided by SKA-LOW and SKA-MID (here, to clarify, we
are referring to both SKA1 and SKA2). For the reasons outlined above
(primarily scattering and $T_{\rm sky}$), SKA-MID will be the
main tool for discovering distant pulsars in the Galactic plane,
whereas SKA-LOW can both go deeper and survey faster at higher
Galactic latitudes. Together, these two surveys will characterize the
spectra and distribution of Galactic radio pulsars in beautiful
detail. They will both discover exotic individual systems which can be
used to test theories of dense matter and gravity. For example the GBT
Drift-Scan survey found PSR~J0337+1715, a unique millisecond pulsar in
a stellar triple system, which promises a very strong constraint on
deviations from the Strong Equivalence Principle of General
Relativity~\citep{rsa+14}.



\subsection{Composite survey with SKA1-LOW and SKA1-MID}\label{sec:skamid}
To consider the usefulness of surveying with SKA1-LOW, we first
compare the number and type of pulsars it will find with those found
by SKA1-MID. We have shown with the help of simulations (see the next
section for details) that all-sky surveys with SKA1-MID find a larger
number of normal pulsars and MSPs than SKA1-LOW. However, with
SKA1-LOW we will find many additional pulsars to those found with
SKA1-MID, thereby the two greatly complement each other.
While young pulsars (ages of $10^3 - 10^7$\,yr) are strongly confined
to their birth places in the Galactic plane (though sources at
$\lesssim 1$\,kpc will still appear to be isotropic), the much older
MSPs (ages of $\gtrsim 10^8$\,yr) will appear across the sky. This
means that an all-sky pulsar survey is critical for finding the best
pulsar clocks for use in direct GW detection and other fundamental
physics experiments. Conversely, a deep Galactic plane search is
needed to find the DNS binaries or the PSR-BH binaries that will
provide the most stringent tests of General Relativity.

To carry out a large area survey with SKA1-LOW would require that a
beamformer and a pulsar search backend be built. In the present
scenario we have considered that we would use the collecting area of
SKA1-LOW out to a radius of 700 m to include a total of 500
stations. Although this is a larger radius than used for SKA1-MID the
significantly lower observing frequency means that the beam size is
about 7 times larger for SKA1-LOW and so to achieve the same survey
speed fewer beams would be required. We consider here a scenario where
we preserve the total number of beams defined in the baseline design
for the SKA1-MID beamformer of about 2048 beams and we split them
across both telescopes with SKA1-LOW having 500 beams and SKA1-MID
having 1500 beams. As the pulsar survey processing cost scales
approximately linearly with the number of beams this transferring of
the beams from SKA1-MID to SKA1-LOW is almost cost neutral. We also
note that in this scenario the decrease in the area to be surveyed
with SKA1-MID would allow for a smaller tied-array beam size.
If one changes the integration times for SKA1-LOW to (say) 1800~s and
SKA1-MID to 2600~s respectively, the relative yield will be different
from the results obtained with a fixed integration time of 600~s for
both telescopes. This optimization can be performed later, when
further results from the ongoing surveys with Parkes, LOFAR and the
GBT are available, enabling an informed judgment. It is clear,
however, that SKA1-LOW should be equipped with searching capabilities.

\section{Simulations}\label{sec:sims}

\subsection{SKA1 Simulations}
To determine how to achieve the maximal possible number of pulsar
discoveries with SKA1 we have performed a number of simulations which
take into account all of the potentially available resources. We then
consider how this might project forward to SKA2. Our simulations
presented here are based on the {\em PSRPOP.py} code~\citep{blrs14}
but are consistent with those obtained with the code used in
\citet{vs10} for LOFAR. The simulations only considered SKA1-LOW and
SKA1-MID, as large area surveys with SKA1-SUR require integration
times that are too long to achieve useful sensitivity and, therefore,
require presently unachievable computational requirements to reveal
the desired binary systems.

We first considered the optimal frequency for pulsar searching with
SKA1-LOW by performing simulations across the entire expected
available band from $100-450$\,MHz assuming sensitivity parameters as
described in the baseline design. We find that, for a 100-MHz
bandwidth, the optimum central frequency is 250\,MHz, where this is a
compromise between pulsar spectral index, sky temperature and
effective collecting area of the log-dipole antennas. Multiple
simulations were then performed for all-sky surveys with SKA1-LOW and
SKA1-MID assuming a fixed integration time of 600\,s per pointing as
described in the baseline design. These show that SKA1-MID would
detect about 9000 normal pulsars and about 1400 MSPs while the numbers
for SKA1-LOW are about 7000 normal pulsars and about 900 MSPs. It
appears, therefore, that SKA1-LOW is less competitive, but it is
important to remember that SKA1-LOW sees a smaller fraction of the sky
and if one looks at the number of pulsars discovered as a function of
dispersion measure, that SKA1-LOW finds more pulsars at low dispersion
measure. This clearly shows that the lower frequencies observed with
SKA1-LOW mean that the dispersion smearing and the scattering in the
ISM along lines of sight to and through the Galactic plane reduce the
number of pulsars that can be detected there, while the superior
collecting area of SKA1-LOW means that it is able to find more pulsars
nearby.

\begin{figure}
  \centering
  \includegraphics[scale=0.45,trim=0mm 0mm 0mm 15mm]{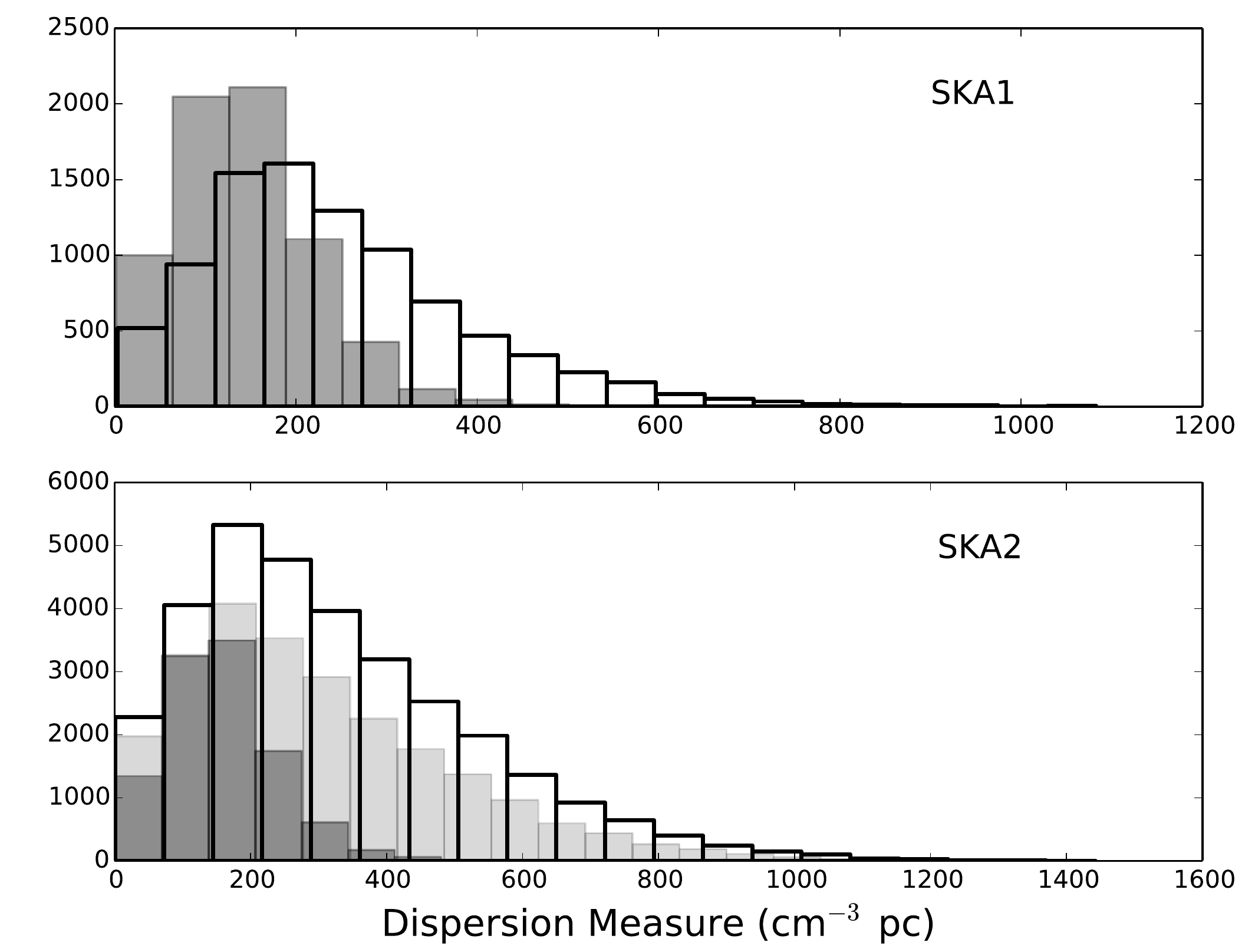}
  \caption{\footnotesize{Histograms showing the pulsar search
      performance of the LOW and MID telescopes in both phases of the
      SKA. In SKA Phase 1 one can see how LOW (Dark bars) performs
      better at low dispersion measures due to the raw sensitivity but
      MID (clear bars) reaches deeper into the Galaxy. With the full
      SKA we consider two options for MID(DISH) of a five- and
      ten-fold improvement in sensitivity, light grey and clear bars
      respectively and also an improved LOW (dark bars). As with SKA1,
      LOW finds a different set of pulsars compared to MID.}}
  \label{fig1}
\end{figure}

\begin{figure}
  \centering
  \includegraphics[scale=0.5, trim=0mm 0mm 0mm 0mm, clip]{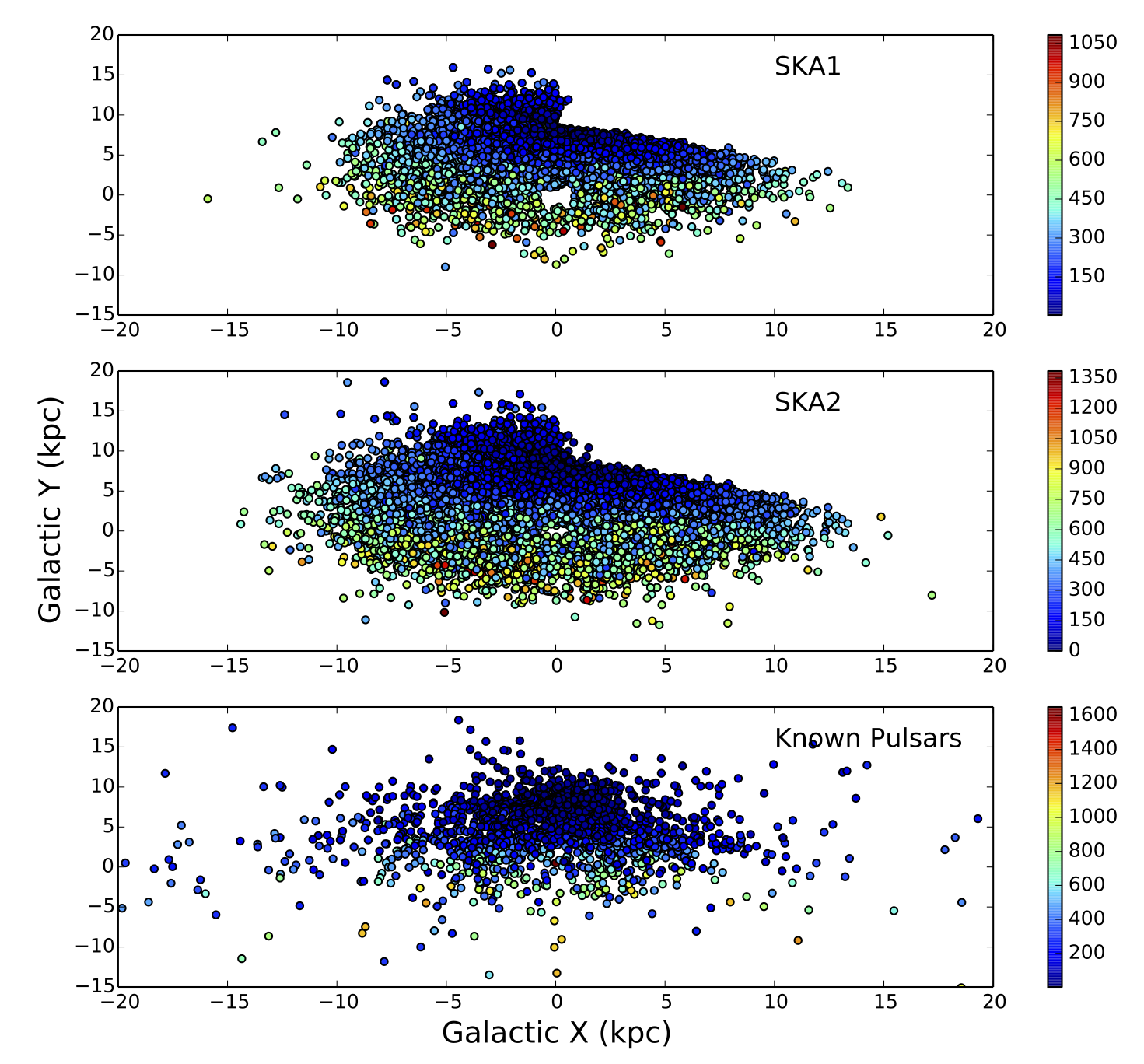}
  \caption{\footnotesize{The number of pulsars expected to be found
      with the SKA and their approximate distribution throughout the
      Galaxy, projected onto the Galactic plane, compared to the known
      pulsar distribution. The colour coding indicates the approximate
      range of dispersion measures of the simulated pulsars that will
      be discovered. We note the dramatic increase in the number of
      pulsars discovered in each phase of the SKA, including a much
      greater sampling of the Galaxy.
      }}
  \label{fig2}
\end{figure}

\textit{The results of these simulations indicate that a composite
  survey, where the complementary regions are covered by the two
  telescopes, provides the best combination to maximise the number of
  pulsars that can be found with SKA1}. We therefore performed
composite survey simulations which used both telescopes to survey
different regions to maximise the pulsar yield and we found that an
optimal survey strategy would 
be to search with SKA1-MID up to Galactic latitudes of about $\pm$10
degrees and the region of the sky in the North where SKA1-LOW cannot
reach. SKA1-LOW would then be used to survey the rest of the sky down
to a Galactic latitude of $\pm$5 degrees. The reason for the overlap
in this region is two-fold: firstly it allows effective
cross-calibration of the two surveys, to ensure that the relative
sensitivities are understood as required for effective modeling of the
population; and secondly because, as we can see from the MSP
simulations, it is apparent that in this region of the sky the two
telescopes are finding different sources, SKA1-LOW faint nearby
objects and SKA1-MID further away objects. This will, therefore,
maximize the return on MSPs, the key target of the survey. In this
scenario we find that we can detect a total of about 10,000 normal
pulsars and
as many as 1500 MSPs.

This highlights the importance of having the beam former and pulsar
search capabilities available for SKA1-LOW in both phases of the
SKA. It also offers up the possibility of an ever-higher pulsar yield,
as with the survey being split across both telescopes, the same survey
can be achieved in the same time with longer integration times,
further enhancing our sensitivity, with the proviso of the highly
accelerated binary systems. 

We have also examined the yield of a preliminary survey with an early
phase SKA1, defined as 50\% the sensitivity of SKA1. For the same
integration times as above, a composite survey would detect about 6000
normal pulsars and as many as 700 MSPs. As expected, doubling the
integration time to increase the sensitivity results in a higher yield
of 7500 normal pulsars and 950 MSPs, but at a much higher, and
impractical, processing cost (a factor of $\gtrsim 10$ times
increase), as per Figure~\ref{fig:cost}.

\subsection{Full SKA Simulations}
We have carried out simulations for the number of pulsars that will be
found with SKA2 in both the LOW and MID (DISH) configurations. In the
case of LOW we have assumed that there is a four-fold sensitivity
increase and that all of that increase can also be applied to the
pulsar search application, when compared to SKA1. In the case of
MID-DISH the nominal improvement in sensitivity is expected to be
about an order of magnitude. However, depending on how those dishes
are distributed and the amount of compute resource that is available
it may not be possible to utilise all of that increase for pulsar
searches. Therefore we consider here two options: the full ten-fold
increase can be used and a five-fold increase in sensitivity. With
SKA-LOW we find a total of 11,000 pulsars including about 1500 MSPs,
while SKA-MID(DISH) will find between 24,000 and 30,000 pulsars, of
which between 2400 and 3000 will be MSPs depending on the exact
improvement in sensitivity. In some regions of the sky \textit{this
  corresponds to detecting the entire population of pulsars that are
  beamed in our direction}. As with SKA1, LOW and MID are highly
complementary with LOW finding the nearby pulsars and MID probing deep
into the Galaxy.

Simulations of pulsar surveys with a mid-frequency aperture array
centered at 750 MHz with 500 MHz of bandwidth show that such a system
would likely be complementary to the LOW and MID surveys already
considered. The survey would detect around 27,000 normal pulsars,
around 6000 of which would not be detected in either the LOW or MID
surveys, and 3000 MSPs, 800 of which would be unique
discoveries. Further analysis will be required on computational costs
and scientific returns to see how the survey will be distributed over
the three antenna types when more detailed specifications are
available.

\section{Targeted Searches}\label{sec:target} 
Targeted searches allow for longer integration times and hence better
sensitivity than wide-area surveys. 
This enables characterisation of specific environments in ways 
unattainable otherwise,
and can elucidate evolutionary links between different types of
neutron stars and their progenitors.

\subsection{Galactic Centre Pulsars}
\label{subsection:Galactic Centre}
The Galactic Centre (GC) is a region of intense interest 
as pulsars discovered here are excellent tools for measurements of the
magnetised ISM in this extreme environment, and could act as unrivaled
probes of the space-time surrounding our nearest supermassive black
hole candidate, Sgr A*
~\citep{lwk+12}. Despite strong evidence for a large neutron star
population in the GC, e.g. $\gtrsim1000$ in the central pc around Sgr
A* \citep{wha12,cl14}, and multi-frequency searches for pulsars, the
number of detections remains low with just 6 active radio pulsars
within $15'$ ($36$ pc) of Sgr A*.
This has primarily been explained by extreme scattering of radio waves
caused by inhomogeneities in the ionised component of the GC ISM
\citep{cl97,lc98}. Scattering, which causes temporal broadening of
pulses, and a corresponding reduction in pulse signal-to-noise ratio
(S/N), can only be mitigated by observing at higher
frequencies. Unfortunately, the steep radio spectra of pulsars
typically prohibits detection above $2$~GHz. Finding the optimum
balance between the effects of pulse scattering in the GC and the
intrinsic luminosity of pulsars has been an on-going problem. It is
therefore clear that the large increase in sensitivity offered by the
proposed collecting areas of SKA1 and the complete SKA 
will greatly help searches in the GC.

The recent detection of an X-ray and radio loud magnetar just 3''
($\sim0.1$~pc) from Sgr A* has both raised hopes for the possibility
of more pulsars in this region, and allowed measurements of the level
of pulse scattering, $\tau_{\rm scatt}$, in this direction
\citep{ken13,mor13,eat13b,sj13,spi14,tkb+14}. While the exact nature
of scattering toward the GC remains uncertain,
these recent measurements suggest that normal slow and some recycled
pulsars might be observable in the GC with SKA1-MID bands 3 and 4
($\tau_{\rm scatt}\sim50$~ms and 7~ms respectively), whereas MSPs will
require band 5 ($\tau_{\rm scatt}\sim4$~ms to $60\,\upmu$s at the
bottom and top of band 5 respectively). For a more detailed discussion
of the prospects for GC pulsar searches and fundamental physics to be
performed therewith we refer the reader to the chapter on observing
radio pulsars in the Galactic Centre~\citep{eat14}.

\subsection{Extragalactic}
\label{subsection:Extragalactic}

Pulsars beyond the disk of the Milky Way are currently known only in
globular clusters and in the Magellanic Clouds, owing to their
intrinsic faintness. While single, non-repeating bursts of apparent
extragalactic origin have recently been detected
\citep{lbm+07,kea+12,tsb+13,spi14}, their astrophysical source remains
unclear. For repetitive, pulsar-like bursts, only much weaker
candidates were found. With the SKA, galaxies in the local group
are within reach using periodicity searches while giant
pulses like those seen from the Crab pulsar can be detected from
galaxies out to well beyond a Mpc.
{\em What is the importance of detecting pulsars in other Galaxies?}
The pulsars likely to be detected will be young with high luminosities
that can be correlated with catalogs of supernova remnants. This will
yield estimates of the star-formation rate and the branching ratio for
supernovae to form spin-driven pulsars as opposed to magnetars and
black holes. Extragalactic pulsars will also provide information about
the magnetoionic media along the line of sight through determination
of the dispersion, scattering and rotation measures.
Unambiguous study of the intergalactic medium in the local group
requires removal of contributions to these measures from the
foreground gas in the Galaxy and gas in the host galaxy. The more
pulsars detected in a galaxy, the more robust this removal will
be. Extragalactic pulsars can be found through blind surveys for both
periodic sources and individual giant pulses. Additional successes
will follow from targeted surveys of individual supernova remnants in
the nearest galaxies. In \S\ref{sec:parameter_space} we discussed the
requirements on sensitivity and FoV. The SKA2-MID sensitivity
will be applied to targeted searches of, for instance, supernova
remnants in nearby galaxies. However, blind surveys over wider fields
such as whole galaxies, will only use the core array. 


Giant pulses from the Crab pulsar serve as a useful prototype for
estimating detection of strong pulses from nearby galaxies. The
strongest pulse observed at $0.43$~GHz in one hour has $S/N_{\rm max}
= 10^4$, even with the system noise dominated by the Crab Nebula.
For objects in other galaxies, the system noise is dominated by
non-nebular contributions, implying that the S/N in this case would
have increased by a factor of about $300$. We can estimate the maximum
distance of detection at a specified signal-to-noise ratio,
$(S/N)_{\rm det}$ as:
\begin{equation}
  D_{\rm max} = \frac{ 1.6\,{\rm Mpc} } { \sqrt{(S/N)_{\rm
        det}/5} } \left (\frac{f_{\rm core} ~ S_{\SKAMid}}{S_{\rm
      Arecibo}} \right)^{1/2} = \frac{ 1.6\,{\rm Mpc} } {
    \sqrt{(S/N)_{\rm det}/5} } \left (\frac{0.4 \times 1630 \, {\rm
      m^2/K}}{{\rm 1150\, m^2/K}} \right)^{1/2}
\end{equation}
where S$_{\rm \SKAMid}$ and $S_{\rm Arecibo}$ are the ratios of
effective area over system temperature for SKA-MID and Arecibo,
respectively (Table 1 in {\tt SKA-TEL-SKO-DD-001}); and $f_{\rm core}$
is the SKA collecting area that can be used for a giant pulse
survey. For $f_{\rm core} = 0.4$, $A_{\rm SKA} / A_{\rm Arecibo}
\approx 1$, the standard one-per-hour pulse seen at Arecibo could be
detected out to $\approx$ 1.2\, Mpc.

In conclusion, SKA will enable not only the discovery of most, if not
all, pulsars in the Milky Way which are beamed towards Earth, but also
allows present-day-survey sensitivities to pulsars in the closest
galaxies. With the single-pulse search techniques, it should be
possible to detect giant pulses from pulsars as distant as the Virgo
Cluster. Studies of the significant numbers of extragalactic pulsars
expected to be detectable by the SKA would allow measurements of the
\emph{intergalactic}, as opposed to the interstellar, medium.

\subsection{Globular Clusters \& High Energy Targets}
\label{subsection:Globular Clusters}
\label{subsection:High Energy Targets}
Targeted searches of globular clusters with SKA1-LOW and SKA1-MID will
utilise the vastly improved sensitivity to discover many new exotic
systems. Pulsars in globular clusters are subject to a much higher
rate of encounters which enables systems to form which would be
impossible in the lower density environment of the Milky Way. For a
detailed discussion of globular cluster search strategies, and the
scienctific applications of these systems, we refer the reader to the
chapter on Globular Clusters~\citep{hes14}.
  
Similarly, targeted searches of unidentified sources found by
high-energy telescopes (such as the LAT on Fermi)
will be employed, as has been done with great success in recent
years~\citep{ray+12}.
These discoveries are mainly highly energetic young pulsars, and
millisecond pulsars, both isolated and in binary systems. The first
group of energetic pulsars is opening a new window to the pulsars'
radio and high-energy emission mechanisms. The second group, of older
neutron stars, is increasing our ability to study neutron star
evolution and is adding numbers to the important population of MSPs,
that are also required to build a Galaxy-size gravitational wave
detector based on these objects (see section \ref{sec:MSP-PTA}). The
SKA will be able to probe high-energy sources that are still
unidentified with greater sensitivity, and more generally it will be
able to search for unknown radio pulsars in any sources found by
telescopes operating across the spectrum. Particularly for very high
energy targets
the wide FoV of the SKA is also a great advantage. Since in many cases
the exact location of the small neutron star within the high-energy
emitting region is unknown, current large single-dish telescopes, with
their small beams, need to do several pointings in order to overcome
these positional uncertainties in the surveys. 
More on the possibility of finding new pulsars in sources discovered
at different wavelengths or in completely different regimes
(e.g. gravitational waves and neutrinos) is described in the chapters
on multi-messenger pulsar science and on the neutron star
population~\citep{lucas14,tkb+14}.

\section{Expected discoveries and their importance}\label{sec:discoveries}

\subsection{Pulsar-Black Hole Binaries and other binaries for GR tests}\label{sec:psrbh}
Since early in the conception of the SKA, the detection and study of
highly relativistic binary pulsar systems has been one of the Key
Science goals~\citep{kbc+04}. Such systems have provided the first
strong evidence for the existence of GWs with the double neutron star
PSR~B1913+16 \citep{tfm79}, the most precise tests of Einstein's
theory of General Relativity (GR) with the double pulsar PSRs
J0737--3039A/B \citep{ksm+06}, and the best tests of alternative
gravity theories with, for example, PSR J0348+0432
\citep{afw+13}. However, in the collection of these remarkable
``gravity labs'' the most prized system has so far eluded detection: a
pulsar - black hole binary (hereafter PSR-BH). The study of a PSR-BH
will allow not only precision tests of GR, but should, for the first
time, allow the properties of the BH to be measured in a model
independent fashion.

{\it The Science prospects from discovery of a PSR-BH system:}
A PSR-BH system will allow us to probe BH properties, as well
providing stringent tests of theories of gravity in general.
If the ``No-hair theorem'' holds, we expect that BHs are remarkably
simple objects described by mass and spin (and perhaps charge)
only. Once the mass and spin are known, other properties like the
quadrupole moment are pre-determined, so that by measuring all
quantities at the same time the no-hair theorem can be tested. For the
same reason, we do not expect a BH to carry a ``scalar charge'', in
contrast to other compact objects like neutron stars where this is
possible. The existence, or not, of a scalar charge results in vastly
different behaviours in the orbital motion of a system involving a BH,
when predictions of GR are compared with alternatives like
tensor-scalar theories. Indeed, it can be argued that a PSR-BH is
probably the best foreseeable probe for testing alternative theories
of gravity~\citep{de96}. Moreover, measuring relativistic spin-orbit
coupling and frame dragging, will allow us to determine the spin of
the BH with high precision. While the mass of the BH can be determined
with high precision, the (unitless) spin parameter should be measured
to be smaller than unity, in order for an event horizon to exist --- a
fact that can, with the SKA, be easily
tested~\citep{kbc+04}. Measuring the effects of classical spin-orbit
coupling will also provide the quadrupole moment of the BH, which can
then be compared to the predictions based on the mass and the
spin. Whether we will be able to measure the spin and quadrupole
moment depends also on the mass of the BH. While \citet{lwk+12} have
shown that it will be difficult to measure the qudrupole moment for
stellar-sized BHs, the BH in the centre of the Milky Way is an ideal
target to attempt this experiment (see e.g. \citealt{eat14}).

\begin{figure}
  \centering
  \includegraphics[scale=0.4,trim=30mm 30mm 0mm 120mm, clip]{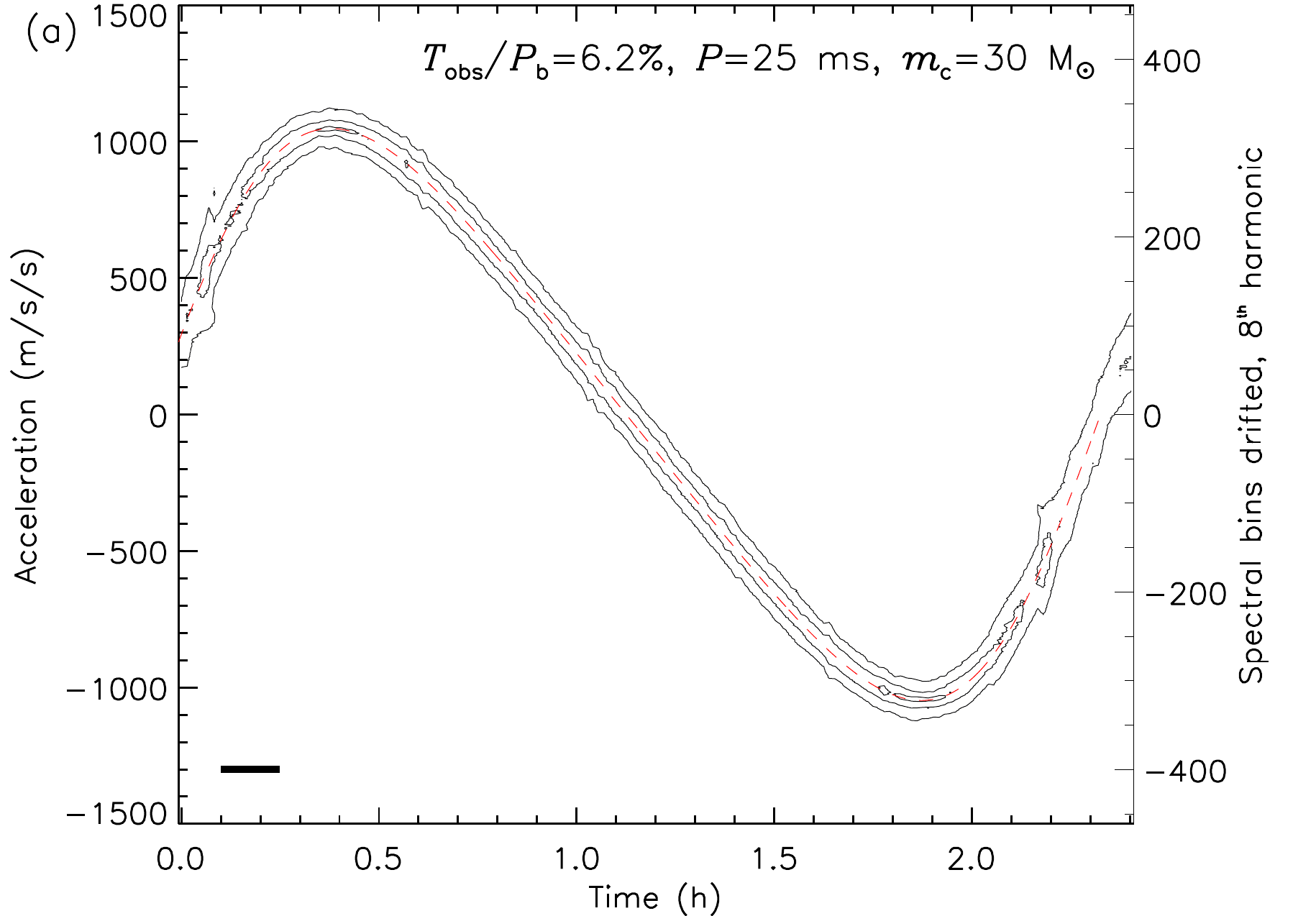}
  \includegraphics[scale=0.4,trim=30mm 30mm 0mm 120mm, clip]{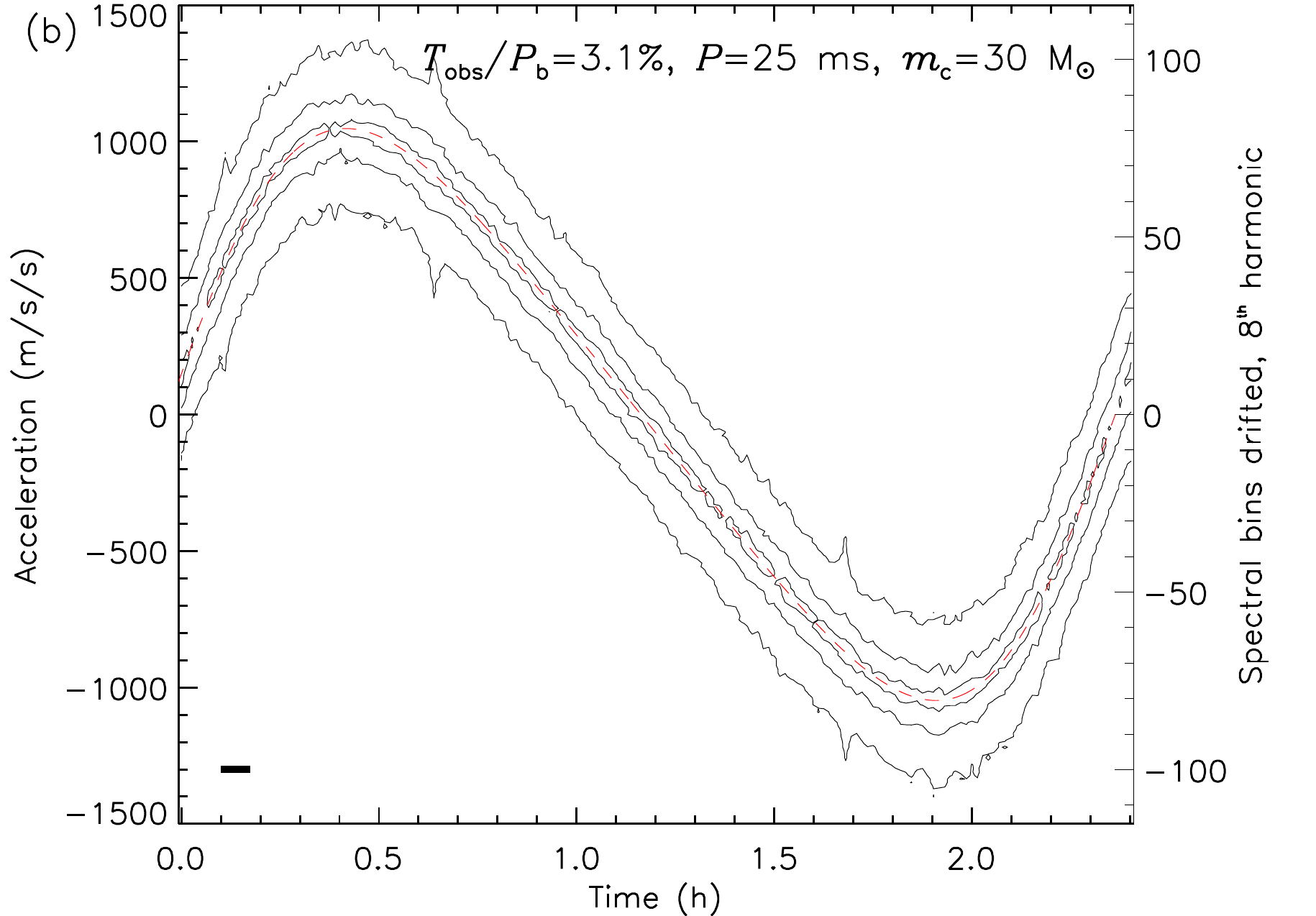}
  \caption{\footnotesize{The effectiveness of acceleration searches in
      the detection of a simulated PSR-BH system.
      While such a system would be ideal for gravity tests, 
      its detection through blind searches would be extremely
      challenging. 
      Here we represent a ``worst-case scenario'', where the pulsar
      has a short spin period of $25$~ms; expected in systems that
      have formed via exchange interactions in dense stellar
      environments, or where the pulsar is the first-born object.
      The black hole mass is at the high end of what is expected for
      stellar mass black holes (30 M$_\odot$), and the orbital period
      is small ($P_{\mathrm{b}} = 2.4$~hr) and has low eccentricity
      ($e = 0.1$). To worsen matters the system is viewed edge-on,
      where the deleterious effects of line-of-sight motion are
      strongest. Contours mark the 30\%, 60\% and 90\% pulsar signal
      recovery levels for constant acceleration searches of data
      integrations which are $\sim6$\% and $\sim3$\% of the orbital
      period (panels (a) and (b) respectively). Because a survey
      observation can start at any phase in the orbit, the starting
      point of the acceleration search has been incremented in steps
      of $50$~s across the entire orbit. The thick black line shows
      the integration length analyzed. The red dashed line shows the
      true acceleration value at the mid-point of integration (Liu et
      al. in prep).}}
  \label{fig_psr_bh}
\end{figure}

{\it The search strategy:} Naturally, these exceptional gravity tests
will first require the detection of a PSR-BH. As outlined above, a
composite survey with SKA-LOW and -MID is proposed to find all pulsars
beaming towards us, with significant inroads being made already with
SKA1.
However, it is fair to say that a PSR-BH might be the most challenging
of all the systems in the ``pulsar zoo'' to detect.
In Figure \ref{fig_psr_bh} the effectiveness of acceleration searches
in the detection of a simulated PSR-BH is shown. Panels (a) and (b)
show contours of 30\%, 60\% and 90\% signal recovery at the orbital
phase at which the integration has started\footnote{\footnotesize Here
  ``signal recovery'' is given by the S/N achieved in an acceleration
  search, normalised by the S/N for an equivalent search of the same
  pulsar, but with no orbital motion.}. For longer integration times
(Panel (a), $\sim6$\% of the orbital period) full pulsar signal
recovery levels are only achieved at a minority of orbital phases
where the acceleration is varying least. By reducing the integration
time, the amount by which the signal is smeared out (due to higher
order effects) in the frequency domain is reduced. Panel (b) shows
acceleration searches on integrations with half the duration
($\sim3$\% of orbital period). Although the raw sensitivity would be
reduced by a factor of $\sqrt{2}$, the recovery level is now $>90$\%
over the majority of the orbit. From sensitivity considerations the
60\% contours in Panel (a) are approximately equivalent to the 90\%
contours in Panel (b), however there is uneven sensitivity coverage
across the entire orbit at 60\% levels in Panel (a). The increased
instantaneous sensitivity, and therefore reduced integration time,
offered by the SKA will undoubtedly improve the chances of detecting a
PSR-BH.

Searches with SKA1 and SKA2 are expected to be performed in real-time
(or pseudo real-time) because of data storage constraints. In searches
for PSR-BH systems, where extreme levels of orbital acceleration might
be observed (see Figure \ref{fig_psr_bh} where the line of sight
acceleration can be greater than $1000~{\rm m\,s}^{-2}$), even
one-dimensional acceleration searches constitute an extreme data
processing task. The degree of computational requirements of
time-domain pulsar acceleration search algorithms has a strong
dependence on the integration time ($\propto T^3$). As shown in Figure
\ref{fig_psr_bh}, acceleration analyses of half length integrations
can result in an equivalent sensitivity to extreme PSR-BH as in longer
integrations, but with 8 times less computational expense. Of course,
it is only because of the superb sensitivity of the SKA that
integration times will be able to be kept to a minimum. In addition,
the reduced integration length also implies that more computational
effort can be spent searching a wider parameter space.

\subsection{MSPs for PTAs and GW searches}
\label{sec:MSP-PTA}
\begin{figure} 
  \centering
  \includegraphics[width=4in]{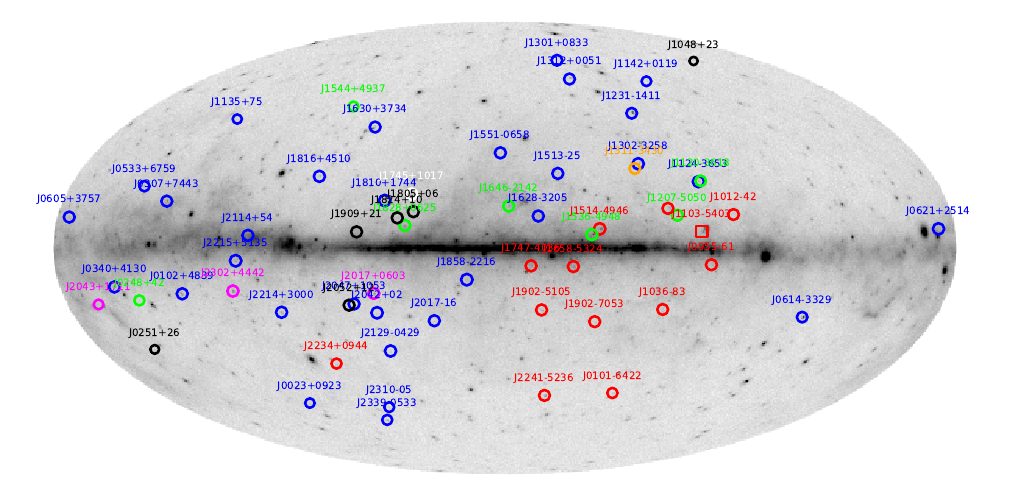}
  \caption{\footnotesize{The initial $\sim$60 newly discovered radio
      MSPs detected via pointed radio searches towards {\em Fermi}
      associated $\gamma$-ray sources.  The nearly isotropic
      distribution of the relatively nearby ($\sim$1$-$3\,kpc) systems
      is obvious.  The colours indicate the discovery telescope: blue
      is the GBT, red is Parkes, black is Arecibo, green is the GMRT,
      magenta is Nan\c cay, and white is Effelsberg.  SKA surveys
      should find {\em thousands} of new MSPs, perhaps 5$-$10\% of
      which will be suitable for high-precision timing work. Searches,
      especially with SKA1, of pulsar-like {\em Fermi} associated
      sources for new MSPs will provide excellent early-science
      opportunities for the SKA. Figure courtesy Paul Ray,
      NRL. \label{fermi_MSPs}}}
\end{figure}

Millisecond pulsars live $10^2-10^3$ times longer than $\sim$1-sec
pulsars and have substantially larger scale heights in the Galaxy.
For these reasons, the majority of the $\sim$200 currently known
Galactic MSPs are {\em local} objects within $\sim$1$-$2\,kpc from the
Sun, and as such, are distributed nearly isotropically on the sky (see
Figure~\ref{fermi_MSPs}). Recent population
studies~\citep{gk13,lorimer13,levin+13} suggest that the Galaxy holds
about 30,000 detectable MSPs in total, thousands of which will be
within reach of SKA pulsar surveys. The MSPs are undoubtedly some of
the most difficult pulsars to detect, due to their rapid spin rates
(demanding fast sampling and high frequency resolution) and their
typical binary nature.
However, their high rotational stability and related timing precision
makes these searches eminently worthwhile, as they can be used for
some of our most important physics experiments, such as the direct
detection of nano-Hz-frequency GWs~\citep{ipta2013}.

\begin{figure}
  \centering
  \includegraphics[width=4in]{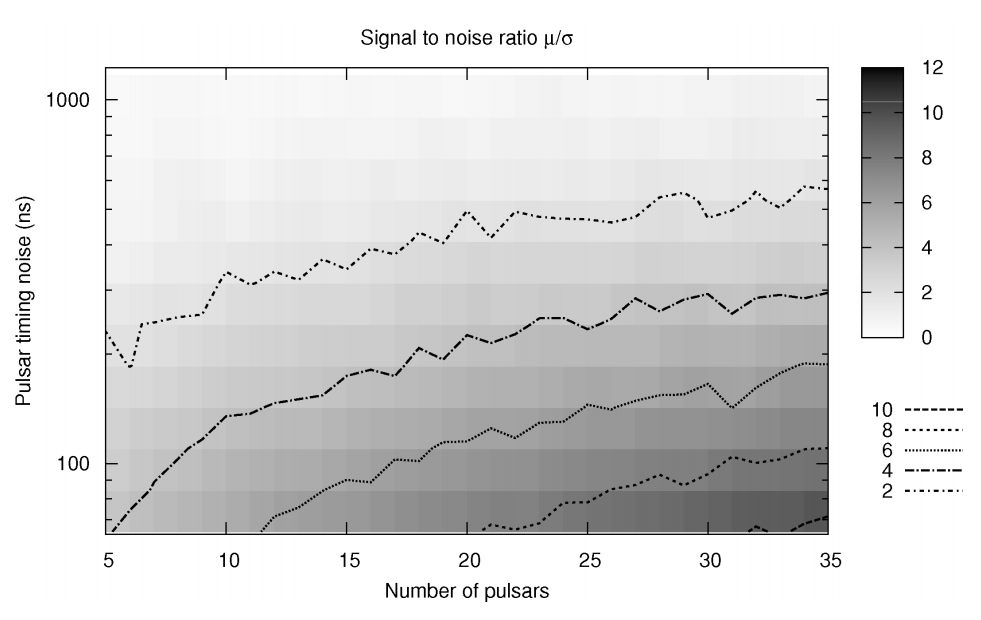}
  \caption{\footnotesize{Density plot (grey scale) and contours of the
      signal-to-noise ratio for GW detection for different
      realisations of pulsar timing residuals, as a function of pulsar
      timing noise and the number of pulsars timed. Here, we have
      assumed 100 data points per pulsars, approximately evenly
      distributed over a period of 7.5 years and a simulated GW
      background with amplitude of
      $10^{−15}$~yr$^{1/2}$~\citep{rutger2009}.
      \label{MSP_PTAs}}}
\end{figure}

Pulsar timing arrays require the very best MSPs, which are selected
based on their flux density, the shapes of their radio pulses (narrow
features are better), their timing stability (unknown until measured),
and their distribution across the sky (a nearly isotropic distribution
is close to ideal for the detection of a stochastic GW background).
SKA all-sky surveys will find thousands of new MSPs, but only a
relatively small fraction --- perhaps 5$-$10\% --- will be of
sufficient quality to include in an SKA-based PTA
(e.g. Figure~\ref{MSP_PTAs}). Finding these rare pulsars is crucial as
recent work has shown that GW sensitivities are {\em directly}
proportional to the number of `good' pulsars being timed
\citep{siemens+13}. In addition, these ultra-stable MSPs will provide
spectacular `secondary' science such as high-precision pulsar masses
which will constrain the Equation of State of nuclear
matter~\citep{dprrh10}.

\subsection{Testing the neutron star equation of state}
\label{sec:NS-EoS}
\begin{figure}
  \centering
  \includegraphics[width=8cm,trim=0mm 50mm 0mm 30mm, clip]{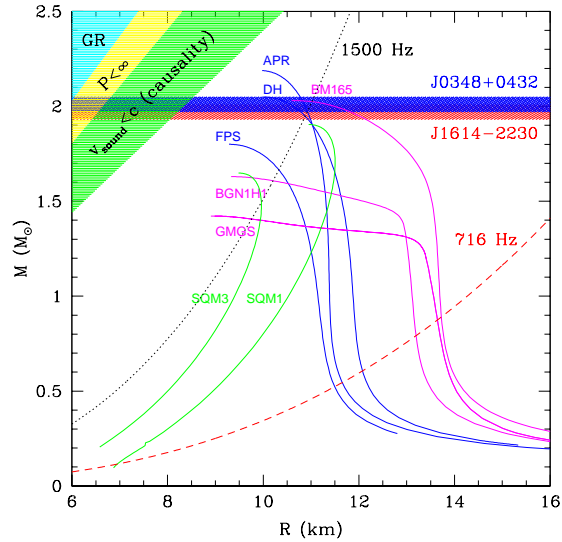}
  \caption{\footnotesize{\label{fig:NS_EOS} Different equations of
      state are shown, with regions in the neutron star mass versus
      radius parameter space which are ruled out or as yet allowed
      highlighted. Figure adapted from \citet{dprrh10}. For further
      details we refer the reader to the chapter on probing the NS
      interior and the cold dense matter equation of
      state~\citep{anna14}.}}
\end{figure}

Together with precise mass determinations in binary systems (see
e.g. \citealt{dprrh10,afw+13}), pulsar spin rates are among the most
accurately known observables which could
constrain the neutron star EoS. 
Pulsars can only be spun up to a limiting period, $P_{\mathrm{\rm
    sh}}$, below which the star, due to centrifugal forces, becomes
unstable to {\it mass shedding} at its equator.
The binary MSP J1748$-$2446ad \citep{hrs+06}, is the neutron star with
the shortest known rotational period, $P_{\rm min}=1.396$
ms\footnote{An 0.89-ms period for the X-ray source XTE
  J1739$-$285~\citep{k++07} still awaits confirmation.},
implying that the radius, if it were a canonical
1.4~$\mathrm{M}_\odot$ neutron star, should be smaller than $\sim 15$
km \citep{hrs+06}.
Irrespective of the proposed EoSs, the period $P_{\rm min}$ of
J1748$-$2446ad must be longer than $P_{\rm sh}.$ 
The value of $P_{\rm sh}$ strongly depends on the
EoS~\citep{pk94,sf95}, with the observed $P_{\mathrm{min}}$ very close
to the mass-shedding limit for the most `stiff' EoSs (see
Figure~\ref{fig:NS_EOS}). However, the re-acceleration of a neutron
star with an initial mass of $1.4$~$\mathrm{M}_\odot$ could in
principle proceed down to limiting spin periods as fast as $0.6$~ms
for most `soft' EoSs \citep{cst94}.
As a consequence, even the discovery of one sub-ms pulsar, with a
rotational frequency well above $1$~kHz, would lead to the rejection
of a wide class of EoSs and provide crucial information about the
behaviour of matter at supra-nuclear densities.

Although few recent pulsar surveys 
could have detected sub-millisecond radio pulsations, a strong bias
exists against detecting fast spinning MSPs as a result of the
observed preference of these neutron stars to be hosted in eclipsing
binaries (9 cases out of a total of 11 binary MSPs with $P<2$ ms,
including PSR J1748$-$2446ad). In these systems matter released by the
companion engulfs the system and obscures the radio pulsations for a
large fraction of the orbital period and, occasionally, for the entire
orbit, particularly at lower radio frequencies. Also, a pulsar like
J1748$-$2446ad is too faint (about 80 $\upmu$Jy at 1.95 GHz) to be
detectable by most of the past and/or ongoing large scale
surveys. Most of these difficulties will be overcome, or strongly
mitigated, by a deep all-sky survey at a frequency of about $2$~GHz
with SKA1-MID. Provided sub-ms pulsars exist, and they mostly reside
in eclipsing binaries, this experiment will provide an unprecedented
opportunity to uncover these objects.

\section{Importance of follow-up timing and strategies}\label{sec:followup}
When new pulsars are discovered their rotational, astrometric and
binary properties (in case of a companion), need to be determined in
order to be able to acquire a stable timing model that can be used to
predict the arrival times of the pulses. It is this which enables the
many high precision measurements possible using pulsars. The first
approximation of a timing model requires very regular observations,
starting from daily monitoring to gradually increased spacing between
observations. In order to separate the position of the pulsar from the
basic spin parameters, at least half a year of timing is necessary
(although positional determination using VLBI methods, where possible,
can remove this requirement). Pulsars in binary orbits need a
high-cadence sampling of the binary orbit to measure the first-order
binary parameters. In case of relativistic binaries, the
post-Keplerian parameters can usually only be measured after multiple
years of timing.

For all pulsars, multi-band monitoring is required to mitigate ISM
effects on the signal. In general, lower frequencies are better for
measuring dispersion measure variations and scattering delays, and
therefore we expect most pulsars will require monitoring at the lower
bands of SKA-MID. In addition, SKA-LOW can also be used for follow-up
timing for the relatively brighter and/or more nearby
pulsars. However, in some cases higher frequencies (around $3$~GHz)
will be essential to get a better coverage of binary orbits.
As described above, the fastest-spinning pulsars are likely to be
found in eclipsing binaries. Due to the fact that the effect on the
higher-frequency part of the signal is smaller, the pulses can be
detected for a larger fraction of the orbit at those frequencies and
makes determining and monitoring the binary parameters possible or
easier.

We aim for regular follow-up for all known and newly discovered
pulsars visible in the SKA sky. Even the slow pulsars need to be
covered in the SKA timing programme on a regular basis. The 45-year
timing programme at Jodrell Bank, where up to 800 pulsars have been
observed, has led to unexpected results on various timescales for all
kinds of pulsars~\citep{lhk+10,elsk11,lgw+13} showing that different
types of pulsars require different strategies in their follow-up and
long-term timing observations. 

{\it Pulsars that show timing noise or rotational irregularities:}
These are typically young pulsars, or high-magnetic field pulsars that
show timing irregularities, mode changes on various timescales or
glitches. The required cadence for observing depends on the relevant
variability time scales and varies from once a week to once a
month. Many pulsars exhibit glitches, sudden (probably instantaneous)
increases in the rotation rate of a pulsar that happen on different
timescales and on an irregular basis~\citep{elsk11}. They are thought
to be the result of changes in the interior of the neutron star where
angular momentum is transferred from the interior to the
crust. Therefore closely monitoring the recovery in the spin
parameters of the pulsar after a glitch can provide us with
information of the glitch mechanism itself, and the equation-of-state
inside the neutron star (for more details see \citealt{anna14}). {\it
  Intermittent and mode-changing pulsars:} these sources show
differences in spin-down rates on various timescales
\citep{ksm+06,lhk+10}.  An observing cadence matching the time scale
of the variability is required. {\it Rotational RAdio Transients
  (RRATs):} These are neutron stars that only emit detectable single
pulses~\citep{mclaughlin2006,km11} on a very irregular basis
and therefore require more integrated observing time and/or a higher
cadence per source to get to a stable timing solution.

{\it Binary MSPs:} observations at least once a month, and occasional
high-cadence (or full-orbit, when orbital periods are less than about
a day) orbital sampling are required to measure the orbital
parameters. GR tests need both dense orbital coverage to measure the
binary parameters accurately, as well as a long-term timing programme
to detect and monitor secular changes in binary orbits and astrometric
parameters, and to disentangle those. {\it Exotic binary MSPs:} the
SKA promises to uncover many exotic systems, e.g. when
PSR-BH systems are found we might expect that the post-Keplerian
effects on the orbit could become more complex than in the currently
known systems that are used for GR tests. Second, or higher order,
post-Newtonian effects would then become important requiring long-term
and high-cadence observations. In this scenario it is necessary to
obtain a high-cadence coverage of the PSR-BH binary orbit (or
full-orbit observations, depending on the orbital period) as well as
daily monitoring of the system to ensure coherence is maintained. It
is likely that other exotic systems like triple systems will be found,
especially in globular cluster searches, as multi-body systems are
more likely to be found there. It has been shown~\citep{rsa+14} that
daily monitoring is required to maintain coherence for those systems.

{\it PTA pulsars}: one of the main goals of long-term timing is to
directly detect low-frequency GWs using a pulsar timing array. A set
of stable MSPs is used as a Galaxy-scale GW detector which depends on
having long-term timing observations for pulsars distributed across
the sky. There are different classes of GWs which may each require a
slightly different observing strategy (for more information, see the
chapter on ``Gravitational wave astronomy with the
SKA'',~\citealt{gemma14}). Overall, given the large number of MSPs
that the SKA will discover (see above), for GW work the requirements
are: (1) observe as many MSPs as possible on a regular basis with
multi-frequency coverage; and (2) identify the most stable pulsars by
long-term monitoring and determining the red-noise contamination in
their timing residuals. As most PTA pulsars are in binaries, they have
the same requirements as those listed above for the binary pulsars.

Follow-up pulsar timing also supports multi-wavelength observations by
providing precise ephemerides. Over a hundred pulsars are now known to
emit pulses in the X$-$ray and/or $\gamma -$ray regimes. To fold their
photons correctly, in many cases the $\gamma -$ray and sometimes
X$-$ray communities are dependent on up-to-date timing ephemerides
based on radio observations. Also, multiwavelength counterpart
observations can be used to independently constrain distances, and
study the pulsar emission mechanism. Thus the short-term follow-up
timing of all newly discovered sources requires a short-spacing
campaign to make an initial coherent timing solution. Depending on the
parameters and the complexity of the source (timing irregularities,
binary parameters, etc) the longer-term timing cadence can be expected
to be somewhere between daily and monthly, with full-orbit campaigns
on a regular basis for the most important or complex systems.

\bigskip

{\it In this chapter, we have demonstrated that significant
  achievements can be made in pulsar searching in the early science
  phase of SKA1 with a highly concentrated core. In contrast, in order
  to achieve the highest precision in pulsar timing, the gain in using
  SKA2 is enormous when compared to timing observations enabled with
  SKA1. It is this gain in timing precision for selected KSP objects
  which will ultimately require the sensitivity of SKA2.}

\setlength{\bibsep}{0pt}

\end{document}